\documentclass[10pt, twocolumn]{article}
\usepackage{graphicx} 
\usepackage{lipsum} 
\usepackage{cuted}

\usepackage{lettrine}

\usepackage{hyperref}
\usepackage{amsmath}

\usepackage[left=1.5cm, top=2cm, bottom=2cm, right=1.5cm]{geometry} 

\usepackage{hyperref} 

\newcommand{\Fig}[1]{Figure~#1}
\newcommand{\Tab}[1]{Table~#1}

\usepackage{multirow}
\usepackage{multicol, booktabs}
\usepackage{adjustbox}

\newcommand{\lFreqEnergy}[1]{Frequency and energy distribution associated with each IM using #1. The boxplot depicts the corresponding results among all the studied participants for the resting-state, motor, and gambling fMRI experiments. The shadowed area highlights the modes whose frequency appears within the neurophysiological band.}

\title{\textbf{Multiscale Functional Connectivity: Exploring the brain functional connectivity at different timescales}}

\author{Manuel Morante, Kristian Frølich, and Naveed ur Rehman\\ \footnotesize{Department of Electrical and Computer Engineering of Aarhus University,} \\ \footnotesize{Aarhus, 8200, Denmark (e-mail: morante@ece.au.dk).}}
\date{}

\begin{document}

\maketitle

\begin{strip}
    \centering
    \begin{minipage}{.9\textwidth}
        \begin{abstract}
            Human brains exhibit highly organized multiscale neurophysiological dynamics. Comprehending those dynamic changes and the neuronal networks involved is critical for understanding how the brain functions in health and disease. Functional Magnetic Resonance Imaging (fMRI) is a prevalent neuroimaging technique for studying these complex interactions. However, analyzing fMRI data poses several challenges. Most approaches for analyzing Functional Connectivity (FC) still rely on preprocessing or conventional methods, often built upon oversimplified assumptions. On top of that, those approaches often ignore frequency-related information despite evidence showing that fMRI data contain rich information that spans multiple timescales. This study introduces a novel methodology, Multiscale Functional Connectivity (MFC), to analyze fMRI data by decomposing the data into their intrinsic oscillatory component, allowing us to separate the neurophysiological activation patterns at multiple timescales while separating them from other interfering components. Additionally, the proposed approach accounts for the nonlinear, nonstationary, and multivariate nature of fMRI while allowing for the particularities of each individual in a data-driven way. We evaluated the performance of our proposed methodology using three fMRI experiments. Our results demonstrate that our novel approach effectively separates the fMRI data into different timescales while identifying highly reliable functional connectivity patterns across individuals.
        \end{abstract}
        \hspace{2cm}
    \end{minipage}
\end{strip}

\section{Introduction}
\label{sec:introduction}
\lettrine{T}{he} brain is a complex system that exhibits highly organized neurophysiological interaction at various time scales \cite{ZalTim_2014, PreDyn_2017, BolTap_2020}. As a result, brain activity displays highly complex spatiotemporal dynamic \cite{BolTap_2020, LurQue_2020}, and their knowledge is crucial to enhancing our understanding of the brain's functioning and cognition \cite{FaiFun_2009, PolHan_2011, FriFun_2011}.

Researchers have used a variety of neuroimaging techniques to study brain activity. Among all those techniques, functional Magnetic Resonance Imaging (fMRI) stands as one of the most popular for both neurophysiological, e.g.,~\cite{OzcNat_2023, BraTas_2022, ChaEmo_2020}, and clinical research, e.g.,~\cite{PenAbn_2021}. Unlike other alternative noninvasive neuroimaging techniques, such as Electroencephalography (EEG), e.g., \cite{SinTre_2023}, or Magnetoencephalography (MEG) \cite{KosEvo_2021}, fMRI records brain activity indirectly by measuring the variation in the oxygenation levels on small volumes of tissue, referred to as voxels~\cite{PowFun_2011}, due to the metabolic oxygen consumption of the neurons, which is a process referred to as Bold Oxygenation Level-Dependent (BOLD) contrast~\cite{PolHan_2011}. In addition to the brain-induced activity, the fMRI data contains a mixture of several interfering signals, such as movement, respiratory, and cardiac pulsations \cite{BiaSou_2009}.

During the last two decades, fMRI research has increasingly focused on exploring the brain's functional connectivity (FC) and the dynamic interactions among neuronal networks, as envisioned by Friston back in 2011~\cite{FriFun_2011}. This shift reflects a broader recognition of the brain's complex organization beyond localized activations in response to stimuli or tasks \cite{FriFun_2011, CalMul_2012,  BolTap_2020}. These conventional approaches focus on unveiling the activation areas and their corresponding activation patterns in response to stimuli, often referred to as sources \cite{FaiFun_2009, PowFun_2011, MorBli_2021}.

Meanwhile, in light of the success of Machine Learning and Artificial Intelligence in a wide range of complicated problems, current approaches for studying FC often leverage machine or deep learning-based models. These models have shown a great potential to uncover hidden interactions among brain networks that are not directly observable from the data \cite{KimRep_2021}. Yet, these approaches still depend on some fundamental preprocessing steps or conventional fMRI analysis techniques, such as Independent Component Analysis (ICA) \cite{CalMul_2012}, which remain critical for their application in practice \cite{PerOpt_2020, BolTap_2020}. These preliminary steps, however, present several limitations, which hinder the realization of the full potential of those more advanced approaches.

For instance, conventional motion correction approaches rely on the parametric rigid-body model, which assumes that the brain only undergoes rigid motion. These approaches provide a reasonable estimation for main motion effects. Nonetheless, these motion components are richer than this simple model, as they can carry more information regarding certain behavioral and physiological aspects of the participants, such as arousal \cite{GuAro_2019}, and even age \cite{BolAgi_2020}. Furthermore, these simple models overlook other more subtle sources of motion, such as respiratory or cardiac pulsations. These additional motion components are often challenging to identify since they appear highly structured, individual dependent, and even mimic other brain sources \cite{CheRes_2020}.

Temporal static filtering constitutes another prevalent preprocessing step that aims to remove portions of the frequency spectrum from the acquired fMRI data with no relevant neurophysiological information \cite{PolHan_2011, LurQue_2020}. However, this step typically involves a predefined selection of filter parameters, such as the number of components, cut-off frequencies, and corresponding bandwidths, which are hard to estimate in practice, among other limitations, see \cite{YueInt_2019, BolTap_2020}. That said, using the signal's frequency information to extract its neurophysiological components and remove unwanted interfering components holds promise and has already been investigated in multiple studies. For instance, \cite{YueInt_2019} have recently explored the idea of decomposing the fMRI signal into their inherent oscillatory components. However, \cite{YueInt_2019} performed their study voxel-wise, ignoring the multivariate nature of fMRI data, and they still used a band-pass filter to remove interfering components. Consequently, finding an adequate filtering process that balances preserving relevant information and removing noisy interfering components remains challenging in fMRI research.

In addition to these preprocessing steps, many approaches for FC analysis still utilize some conventional fMRI source separation techniques as part of their preprocessing pipelines \cite{LurQue_2020, MorInf_2020}. One of the most relevant approaches is ICA \cite{CalMul_2012}, which plays a crucial role in many analysis pipelines by (a) separating neuronal networks of interest, (b) separating interfering components, e.g., cardiac or motion artifacts, from neurophysiologically relevant sources \cite{XuDen_2014}, and (c) as a data-driven way to parcellate the brain into relevant brain sources. Despite its wide use in fMRI data analysis, ICA assumes that the fMRI data is --from a statistical point of view-- independent and stationary\footnote{Formally, ICA is a statistical-based method that assumes that the data of interest can be described as a linear combination of a set of maximally independent sources. The assumption of statistical independence also implies that the data are weekly stationary \cite{Sergios_2020}.}, an assumption often overlooked in practice and may not correspond with the natural brain behavior \cite{MorInf_2020}.

Nevertheless, although alternative approaches rely on similar assumptions, there is compelling evidence of the brain's nonlinear and nonstationary nature. Authors in \cite{GuaPro_2020} explicitly studied the stationary and nonstationary behavior of different brain areas, showing that the brain dynamics are nonstationary and nonlinear. Furthermore, \cite{ZalTim_2014} and \cite{PreDyn_2017} reported consistent evidence supporting brain nonstationary behavior. As emphasized in \cite{GuaPro_2020}, researchers should consider nonlinear and nonstationary effects when studying brain FC.

Consequently, it becomes evident that there is a need for more advanced and adaptive preprocessing methods capable of incorporating the nonlinear and nonstationary nature of fMRI data \cite{BolTap_2020, GuaPro_2020, CorFre_2001}. By separating neuronal brain activity from other interfering components and unraveling neuronal activity across multiple frequency bands, we address these challenges. Our proposed approach, referred to as Multiscale Functional Connectivity (MFC), allows measuring the brain's FC at different timescales without any static preprocessing or source separation, with the aid of the Multivariate Mode Decomposition (MMD) \cite{HuaEmp_1998, RehMul_2009, RehMul_2019} model. In addition, it is fully data-driven, and it automatically tunes the parameters of the relevant frequency bands based on the input data.

In summary, we propose an alternative way to extract FC information from fMRI data, bypassing some of the limitations of conventional preprocessing steps. In addition to aligning with the expected natural complexities of the fMRI data, our proposed approach can separate interfering components, such as structural noise from heart and respiration components, and simultaneously offers new insights into how the brain networks operate at different timescales. Finally, we evaluated the performance of our proposed methodology using three distinct fMRI studies. Our results demonstrate that our novel approach effectively separates the fMRI data into different timescales while identifying FC patterns with high reproducibility across individuals, and we further extended our knowledge of how the FC for these three experiments spans among different timescales.

\section{Methods}
\label{sec:Method}

\subsection{Background: Multivariate Mode Decomposition}

Formally, Multivariate Mode Decomposition (MMD) is a Signal Processing model that assumes that a multivariate signal of interest accepts a representation as a linear combination of a set of a particular family of amplitude- and frequency-modulated (AM-FM) functions, with a well-defined instantaneous frequency at any given time instance among all the channels \cite{HuaEmp_1998, RehMul_2019}, which are formally defined as $c_{k}(t) = a_{k}(t)\cos(\phi_{k}(t))$, where $a_{k}(t)$ and $\phi_{k}(t)$ are the amplitude and instantaneous frequency functions associated with the $k$-th component. In this way, the MMD model decomposes a multivariate signal, say $\boldsymbol{x}(t)=[x_1(t), x_2(t),\ldots, x_R(t)]^{\mathsf{T}}$, where $R$ is the total number of channels, into a linear combination of intrinsic oscillatory components, referred to as Intrinsic Modes (IMs), as follows:
\begin{equation}
    \boldsymbol{x}(t)=\sum_{k=1}^{K}\boldsymbol{c}_{k}(t),
    \label{eq:MMD}
\end{equation}
where $\boldsymbol{c}_{k}$ is the $k$-th multivariate component and $K$ is the total number of components \cite{RehMul_2019}. In other words, these IM functions behave similarly to harmonics that remain relatively close to a particular frequency. Yet they are flexible enough to accommodate fluctuations in both amplitude and frequency in a data-driven way \cite{DraVar_2014}. 

Unlike other similar alternative studies, e.g., \cite{YueInt_2019}, MMD inherently exploits the multivariate nature of fMRI data. The multivariate nature of MMD allows each voxel or region of interest (ROI) within each specific IM to exhibit a distinct functional activation pattern while sharing a common oscillatory behavior, i.e., a common central frequency. This flexibility accommodates a wide range of response changes. This behavior also contrasts with conventional approaches, such as ICA, where each voxel/ROI activity is decomposed as a linear combination of a single common time activation pattern \cite{CalMul_2012}.

However, performing MMD constitutes a challenging task, and several methods have been proposed to solve it. Among all the available approaches, in this study, we will focus only on two of the most popular alternatives: Multivariate Empirical Mode Decomposition (MEMD) \cite{RehMul_2009} and Multivariate Variational Mode Decomposition (MVMD) \cite{RehMul_2019}. Nevertheless, although both algorithms aim to achieve the same signal decomposition in a fully data-driven fashion, they are vastly different from an algorithmic perspective.

\paragraph*{MEMD.} MEMD obtains intrinsic modes, $c_k(t)$ in \eqref{eq:MMD}, using an iterative sifting process. The sifting process unfolds by iteratively estimating the local mean of a signal by averaging its maximum and minimum envelope signals. The local mean signal is, by design, a low-frequency signal that is subtracted from the original signal to obtain a candidate for an IM. The drawback of MEMD is that it is highly sensitive to noise and suffers from the mode-mixing problem --a single mode carrying fundamentally different frequencies \cite{EriDat_2023}.

\paragraph*{MVMD.} MVMD\footnote{The implemented code for the MVMD algorithm is openly available on the MATLAB file exchange: \url{https://se.mathworks.com/matlabcentral/fileexchange/72814-multivariate-variational-mode-decomposition-mvmd}} obtains a similar decomposition as given in \eqref{eq:MMD} but unlike MEMD, uses an optimization formulation to estimate all intrinsic modes of a signal simultaneously. The objective is to find an estimate of narrowband functions from the data, where the center frequency of each mode aligns across all the multivariate components of the signal. This renders MVMD more robust to noise and provides a better estimate of the modes \cite{RehMul_2019}.

\subsection{Natural frequencies in fMRI data}
\label{sec:NaturalFreq}

Understanding the frequency organization of the fMRI signal and brain activity dynamics is crucial for MFC. While studies focusing on frequency-related aspects of fMRI are relatively sparse, existing research offers valuable insight into the frequency organization of the fMRI signal and brain dynamics. For instance,~\cite{CorFre_2001} demonstrates that the frequency contribution to the correlation patterns spans several frequency bands. Similarly,~\cite{YueInt_2019} investigated the inherent frequency components across different brain locations --in a voxel-wise fashion--  yielding similar findings. 


Overall, fMRI frequency components comprise a rich spectrum that covers several relevant frequency bands. First, very low-frequency oscillations, lower than 10~mHz \cite{PowFun_2011}, correspond to trends, scanner instabilities, and motion residuals. Neurophysiological activation patterns resulting from neuronal activity appear within the range of 10 to 200 mHz ~\cite{CorFre_2001, YueInt_2019, LurQue_2020} emphasizing the significant contribution of this frequency band to fluctuations related to brain activity, which corresponds with the natural band dominated by the BOLD response.

Additionally, fundamental respiratory oscillations occur around 250 mHz, while the first harmonic of respiration appears around 500~mHz \cite{FraEst_2001}. Contributions from blood vessels and cerebrospinal fluid pulsations fall within 400 to 800~mHz band. Similarly, those high-frequency components exhibited significant structured correlations among different brain areas due to the distinct anatomical distribution of the cerebral blood vessels and ventricles \cite{CorFre_2001}. Similarly, they also pointed out that the cardiac pulsation can spread to lower frequencies due to aliasing, appearing as additional interfering structured components, which complies with the observations in \cite{SooRes_2021}.

\subsection{Proposed approach: Multiscale Functional Connectivity}

\begin{figure*}
    \includegraphics[width=1\textwidth]{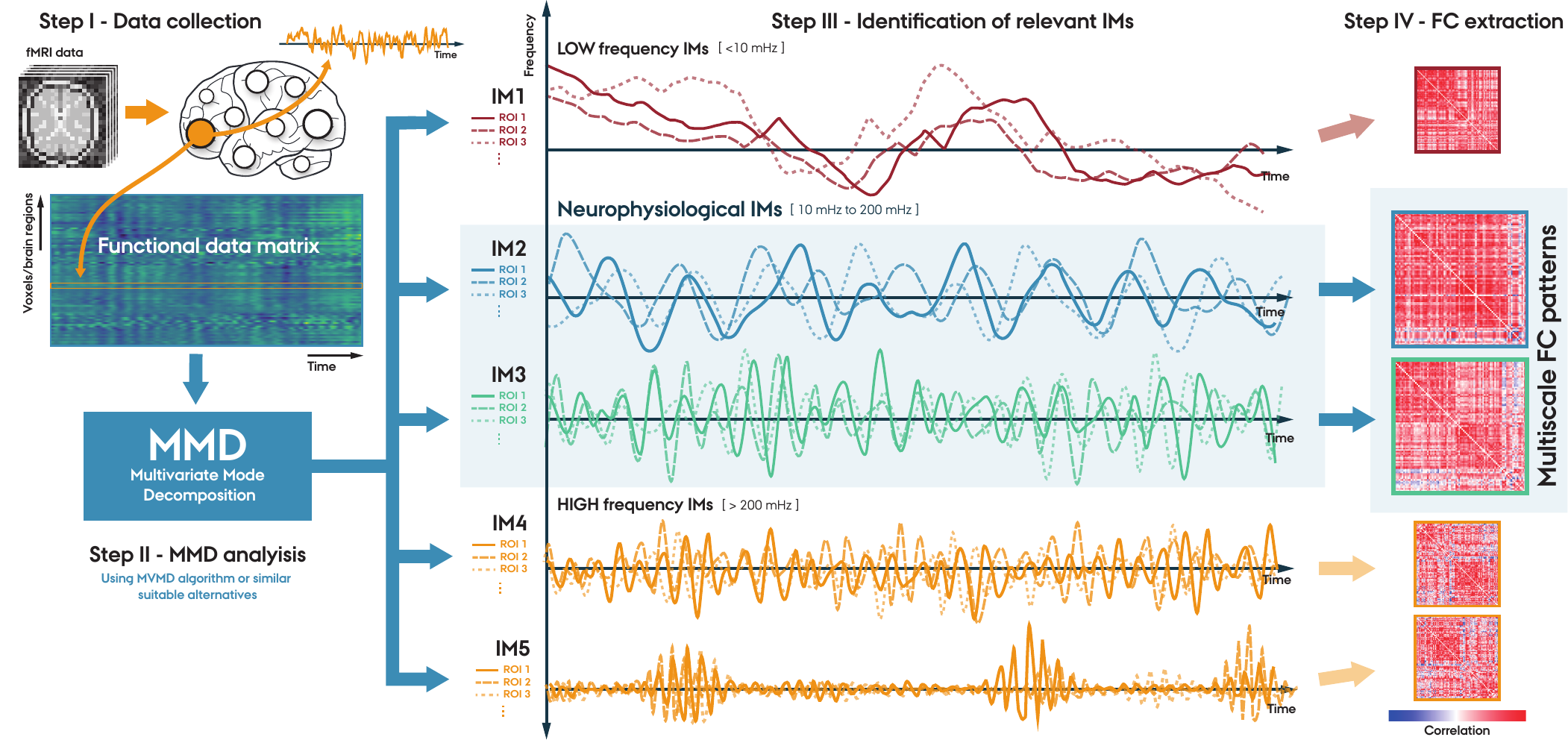}
    \caption{\textbf{Multiscale Functional Connectivity (MFC).} The proposed MFC methodology involves four key steps. (I) Data collection, where fMRI data is aggregated either at a voxel level or through predefined brain atlases;  (II) MMD analysis, which consists of extracting the Intrinsic Modes (IMs) for each individual; (III) Identify the relevant IMs within the neurophysiological frequency band (10-200 mHz); and (IV) FC extraction associated to the IMs of interest.}
    \label{fig:MFC_Methodology}
\end{figure*}

In this study, we propose a novel methodology for analyzing fMRI called Multiscale Functional Connectivity (MFC). \Fig{\ref{fig:MFC_Methodology}} illustrates the steps of this proposed approach.

\begin{itemize}
    \item \textbf{Step I. Data collection.} The first step involves collecting data from each individual. Depending on the specific study, this step can be performed at the voxel level or by using available brain atlases to define some particular ROIs.

    \item \textbf{Step II. MMD analysis.} We perform MMD analysis on the data collected using established algorithms. This analysis obtains the intrinsic oscillatory components (IMs) associated with each inidividual.

    \item \textbf{Step III. Identification of the relevant IMs.} Once the IMs are obtained, we identify the relevant IMs by examining their corresponding central frequencies. Specifically, we focus on the components that fall within the neurophysiological frequency band 10-200 mHz, as discussed in Section \ref{sec:NaturalFreq}.

    \item \textbf{Step IV. Functional Connectivity (FC) extraction.} We can use the obtained multivariate IMs associated with each particular frequency band to uncover the FC at various time scales, providing a complete multiscale FC representation of the fMRI data.

\end{itemize}

While there exist numerous approaches available for estimating the brain's FC \cite{BolTap_2020, LurQue_2020}, in this study we decided --for simplicity-- to use conventional Pearson's correlation coefficient analysis to calculate the correlation between the time activation patterns within each IM. Note that we can do this because, unlike \cite{YueInt_2019}, who performed a voxel-wise study, we unveiled a multivariate representation of each IM. 

On the other hand, this approach differs from static FC analysis, where the raw FC signals are analyzed directly from the ROIs, often after applying static temporal filtering, and from conventional dynamic FC, such as sliding-windows approaches.

\section{Data and preprocessing}

\subsection{Experiment description and fMRI data}
In this study, we investigated three experiments from the WU-Minn Human Connectome Project (HCP) \cite{HCP}. Specifically, we selected the resting state, motor, and gambling experiments from the HCP repository\footnote{Human Connectome Project: \url{https://www.humanconnectome.org/}\label{fn:HCP}}. In each experiment, we randomly selected 100 healthy participants aged 22 to 35 years.

The first experiment was resting-state, where participants were instructed to remain as still as possible during the scan. We chose this experiment because resting-state data is widely used for FC analysis, often providing reliable results.

The motor experiment followed a standard block paradigm, where a visual cue asked the participants to tap their left or right fingers, squeeze their left or right toes, or move their tongue. Each movement block lasted 12 seconds and was preceded by a 3-second visual cue. Additionally, there were extra fixation blocks of 15 seconds each. We chose this simple experiment because the activation patterns and neuronal networks involved are well studied \cite{MorInf_2020}, facilitating the evaluation of the results.

Last but not least, the gambling experiment followed a random block paradigm, where participants tried to guess if a randomly generated number between 1 and 9 was either higher or lower than 5. There are two main reasons why we investigated this additional task-related experiment. Firstly, this experiment has been well studied, making it easier to evaluate the results. The second reason is that, unlike the motor experiment, the gambling experiment's paradigm is unpredictable, i.e., the guesses of the participants cannot be determined a priori. This randomness adds an unpredictable effect to the responses, increasing the variability in the data and allowing for a more robust and consistent analysis. Additionally, we expect the level of arousal and effort for this experiment to be higher than for the other two experiments.

\subsection{Preprocessing}

We obtained the fMRI data directly from the HCP repository\footref{fn:HCP}. The three used datasets were collected using a 3T scanner with a repetition time (TR) of 720 ms. The specific descriptions of the experimental procedures and acquisition parameters are detailed in the HCP imaging protocols\footnote{HCP 3T Imaging Protocol Overview: \url{http://protocols. humanconnectome.org/HCP/3T/imaging-protocols.html}}. In particular, we used the data with minimal preprocessing steps, including motion correction and spatial normalization. Finally, on top of the standard preprocessing pipeline already applied by the HCP \cite{BarFun_2013, HCP}, we further smoothed each brain volume with a 4-mm FWHM Gaussian kernel. 

As discussed in the introduction, we want to emphasize that we do not need any further preprocessing to perform MMD. Unlike \cite{YueInt_2019}, our proposed approach does not require additional preprocessing steps, static temporal filtering or source separation.

\subsection{Selected networks and regions of interest}
For this study, we divided the brain into several regions of interest (ROIs) using the Automated Anatomical Labeling (AAL) atlas \cite{TzoAut_2002}. Although the AAL atlas maps the entire brain, we only analyzed cerebral regions, which resulted in 90 ROIs. Following the recent module-based network organization proposed by~\cite{ParFun_2020}, we grouped these 90 ROIs into seven functional modules. \Tab{\ref{tab:selectedROIs}} contains information regarding the ROIs selected from the AAL and its modules. For the selected 90 ROIs, we extracted the related time series using Nilearn toolbox\footnote{Nilearn: \url{https://nilearn.github.io/stable/index.html}}. Finally, we removed the mean value from each ROI.

\begin{table}[!h]
    \begin{adjustbox}{width=0.5\textwidth, center}
%
\begin{tabular}{clll}
    \hline 
    ROI & Region & \multicolumn{2}{l}{Module}\tabularnewline
    \hline 
    \hline 
    7/8 & Frontal middle & \multirow{7}{*}{FP} & \multirow{7}{*}{Fronto-Parietal}\tabularnewline
    9/10 & Frontal middle orbital &  & \tabularnewline
    11/12 & Frontal inferior opercular &  & \tabularnewline
    13/14 & Frontal inferior triangular &  & \tabularnewline
    15/16 & Frontal inferior orbital &  & \tabularnewline
    59/60 & Parietal superior &  & \tabularnewline
    61/62 & Parietal inferior &  & \tabularnewline
    \hline 
    1/2 & Precentral & \multirow{11}{*}{TP} & \multirow{11}{*}{Temporo-Parietal}\tabularnewline
    17/18 & Rolandic operculum &  & \tabularnewline
    19/20 & Supplementary motor area &  & \tabularnewline
    29/30 & Insula &  & \tabularnewline
    57/58 & Post-central &  & \tabularnewline
    63/64 & Supramarginal &  & \tabularnewline
    69/70 & Paracentral lobule &  & \tabularnewline
    79/80 & Heschl &  & \tabularnewline
    81/82 & Temporal superior &  & \tabularnewline
    83/84 & Temporal pole superior &  & \tabularnewline
    89/90 & Temporal inferior &  & \tabularnewline
    \hline 
    71/72 & Caudate & \multirow{3}{*}{BG} & \multirow{3}{*}{Basal Ganglia}\tabularnewline
    73/74 & Putamen &  & \tabularnewline
    75/76 & Pallidum &  & \tabularnewline
    \hline 
    43/44 & Calcarine & \multirow{7}{*}{Occ} & \multirow{7}{*}{Occipital}\tabularnewline
    45/46 & Cuneus &  & \tabularnewline
    47/48 & Lingual &  & \tabularnewline
    49/50 & Occipital superior &  & \tabularnewline
    51/52 & Occipital middle &  & \tabularnewline
    53/54 & Occipital inferior &  & \tabularnewline
    55/56 & Fusiform &  & \tabularnewline
    \hline 
    3/4 & Frontal superior & \multirow{8}{*}{DMN} & \multirow{8}{*}{Default Mode Network}\tabularnewline
    23/24 & Frontal superior medial &  & \tabularnewline
    31/32 & Cingulum anterior &  & \tabularnewline
    33/34 & Cingulum middle &  & \tabularnewline
    35/36 & Cingulum posterior &  & \tabularnewline
    65/66 & Angular &  & \tabularnewline
    67/68 & Precuneus &  & \tabularnewline
    85/86 & Temporal middle &  & \tabularnewline
    \hline 
    5/6 & Frontal superior orbital & \multirow{8}{*}{Lim} & \multirow{8}{*}{Limbic}\tabularnewline
    21/22 & Olfactory &  & \tabularnewline
    25/26 & Frontal medial orbital &  & \tabularnewline
    27/28 & Rectus &  & \tabularnewline
    37/38 & Hippocampus &  & \tabularnewline
    39/40 & Para-hippocampus &  & \tabularnewline
    41/42 & Amygdala &  & \tabularnewline
    87/88 & Temporal pole middle &  & \tabularnewline
    \hline 
    77/78 & Thalamus & Th & Thalamus\tabularnewline
    \hline 
    \end{tabular}
    \end{adjustbox}
    \caption{Summary of the selected ROIs organized according to the functional modules described by \cite{ParFun_2020}. The numeric label refers to the 90 ROIs associated with the AAL from \cite{TzoAut_2002}, where odd and even numbers correspond to the right and left hemispheres, respectively.}
    \label{tab:selectedROIs}
\end{table}

Additionally, only for the motor experiment, we considered an extra 5 ROIs for the analysis of the time courses associated with the different parts of the motor cortex. For extracting these areas, we used the same motor templates for separating these motor ROIs as, for example, the one implemented by \cite{MorEnh_2021}.

\subsection{Parameter selection}

\subsubsection*{MEMD}
Multivariate Empirical Mode Decomposition (MEMD) requires the specification of two parameters: the number of directions for the projections and the stopping vector. We chose the number of directions for MEMD to be 180, which is twice the value of studied channels, as recommended by \cite{RehMul_2009}, since using a larger number of directions is computationally more expensive, and the improvements on the quality of the obtained intrinsic modes (IMs) will be minimal. The stopping vector contains three different tolerance values. These tolerance values are set to check whether the obtained components comply with the definition of an IM. Particularly, these criteria check for the mean envelope, standard deviation, and sifting difference. These three properties combined guarantee the extraction of meaningful IMs, and optimizing them does not have a critical impact on the performance of the algorithm. Therefore, we selected the default values of the algorithm proposed by \cite{RehMul_2009}, which have been shown to yield good results.

\subsubsection*{MVMD}
\begin{figure*}
    \centering
    \includegraphics[width=1.0\textwidth]{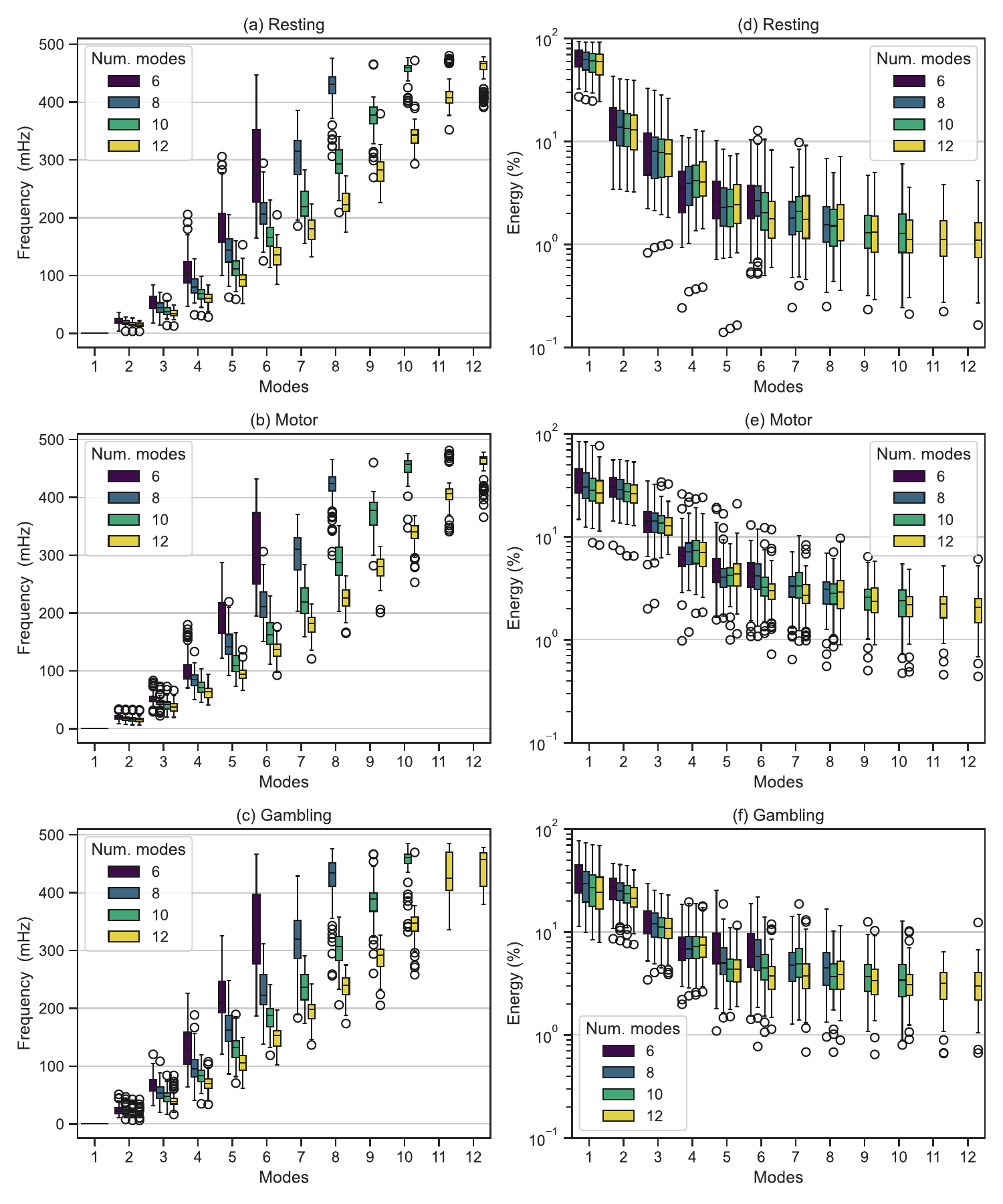}
    \caption{Frequency and energy distribution associated with each mode using MVMD for $\alpha=1000$. The boxplots show the effect of total modes, $K$, for resting-state, motor, and gambling task fMRI experiments.}
    \label{fig:Parameter_modes}
\end{figure*}

Multivariate Variational Mode Decomposition (MVMD) requires two parameters to be manually tuned: the number of modes $K$ and the regularization parameter $\alpha$. The parameter $K$ naturally represents the number of components we expect to observe in the data. On the other hand, the parameter $\alpha$ is a regularization parameter introduced during the optimization that imposes a constraint over the bandwidth of each IM. In this way, a higher value results in intrinsic modes with narrower frequency bands. Since the values of these parameters are unknown a priori, we performed a cross-validation over the 100 participants to identify the optimal values for these parameters. We did this by varying the number of modes, $K$, in the interval of $6-12$ and analyzing the resulting center frequencies and energy distribution. \Fig{\ref{fig:Parameter_modes}} displays the results from this study.

The first column in \Fig{\ref{fig:Parameter_modes}} shows that at least 8 modes were needed to separate the intrinsic modes adequately. For instance, when looking at the case $K=6$, we observed that modes 5 and 6 exhibited a larger variance, indicating that the high-frequency components were not separated correctly. Therefore, $K=8$ constitutes a suitable lower bound for $K$. The second column in \Fig{\ref{fig:Parameter_modes}} shows the relative energy of the modes, where we observed that the energy present in the modes is decreasing for the higher modes. This means using more than 10 modes will contain mostly noise or residual information, whereas high-frequency modes split into more and more components that barely contribute to the signal. For instance, the results for $K=12$, where the modes with the highest frequencies exhibited minimal contributions to the relative energy of the signal. Consequently, after evaluating all these arguments, we decided to set the value to $K=10$, which is big enough to provide enough components and, at the same time, preserves the most information.

\begin{figure*}
    \centering
    \includegraphics[width=1.0\textwidth]{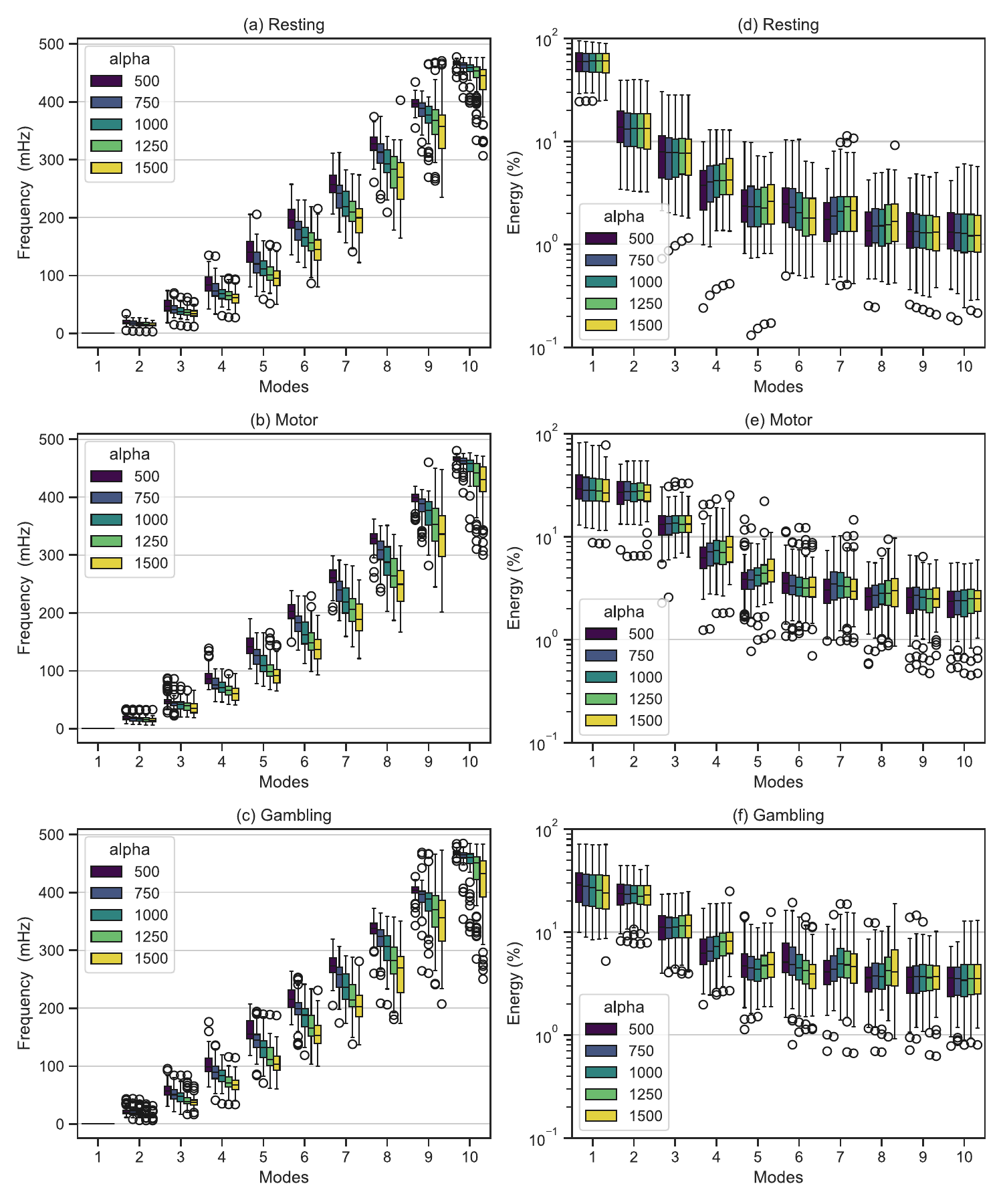}
    \caption{Frequency and energy distribution associated with each mode using MVMD for $K=10$. Boxplots depict the effect of $\alpha$ among modes. The boxplots are for all the studied participants for resting-state, motor, and gambling task fMRI experiments.}
    \label{fig:Parameter_alpha}
\end{figure*}

Regarding the second parameter, $\alpha$, associated with MVMD, we selected this parameter similarly to $K$. For this case, we explored the values of this parameter within the interval $500 - 1500$. The resulting plots from this study can be seen in \Fig{\ref{fig:Parameter_alpha}}. The first column shows that a higher value of $\alpha$ results in more narrow bands for the corresponding IMs within the neurophysiological band of $10-200$~mHz, which is preferable. Looking at the second column in \Fig{\ref{fig:Parameter_alpha}}, we observed minimal changes in the relative energy distribution. This result was expected \cite{RehMul_2019}, as $\alpha$ is a regularization parameter introduced during the optimization. For these reasons, we set the value of $\alpha=1000$.

\section{Experimental results and Discussion}
Through three fMRI experiments, we evaluated the proposed MFC methodology. To this end, we conducted a comprehensive study of two of the most popular MMD algorithms. Then, we analyzed the obtained IMs, including their neurophyisiological relevance, for each one of the study algorithms. In addition, we examined the FC associated to the IMs and the reproducibility of the results among participants.

\subsection{Time activation patterns from the motor-task experiment}
We selected the primary visual cortex (ROI 43 and 44) for the visual responses. In addition, we further divided the ROI from the motor cortex into five additional motor-related ROIs for the different motor areas, corresponding to the right/left hands, right/left feet, and tongue \cite{MorInf_2020}. Note that the ROIs related to the motor tasks have been well documented and studied \cite{TurSma_2018, BarFun_2013, YeoOrg_2011}.

\begin{figure}
    \centering
    \includegraphics[width=1.0\columnwidth]{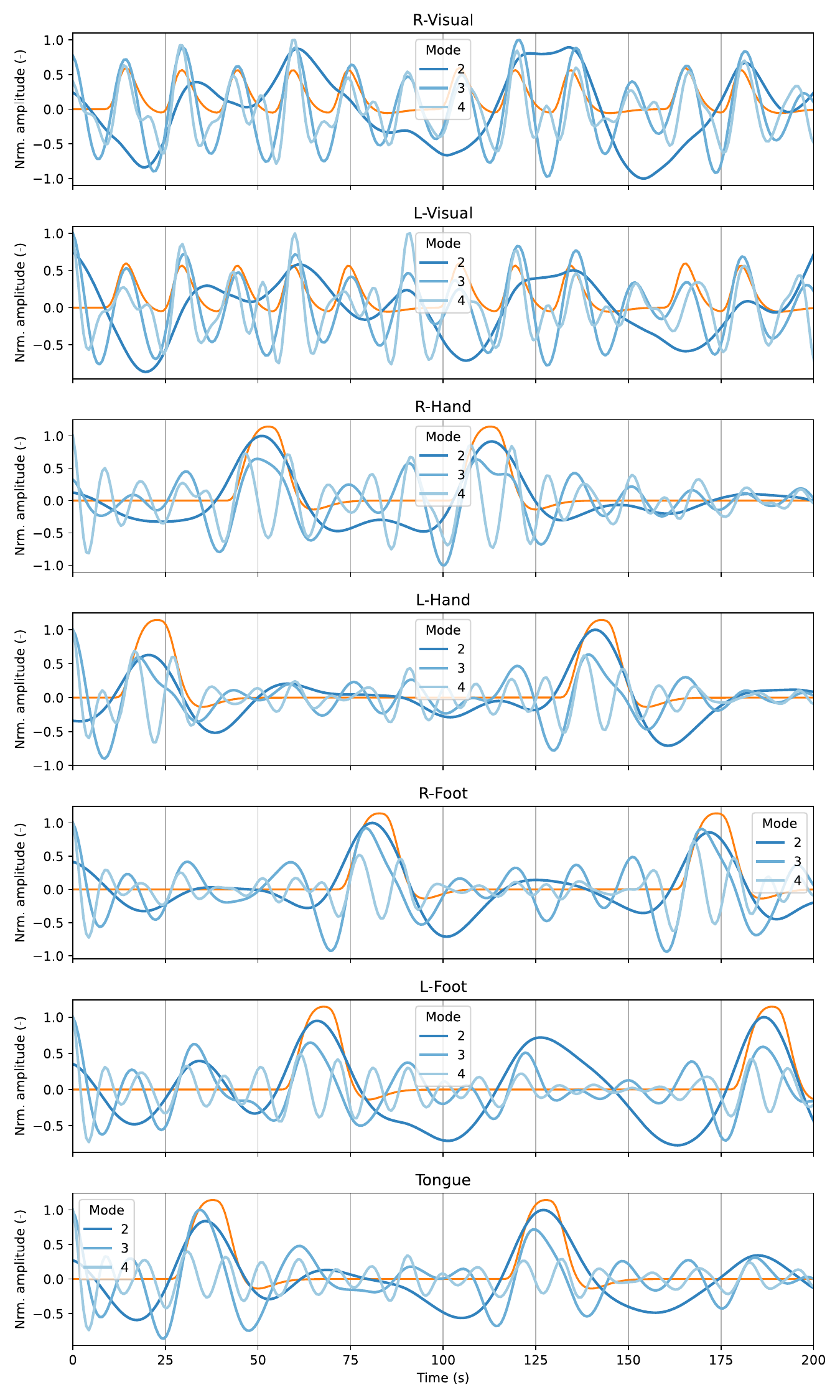}
    \caption{Average time courses among all the studied participants associated with the first neurophysiological modes (2 and 3) from the results of MVMD (blue lines), and the canonical task-related component (orange line) for the most relevant ROIs associated with the motor experiment.}
    \label{fig:Motor-TC}
\end{figure}

\Fig{\ref{fig:Motor-TC}} illustrates the average time courses for the relevant ROIs associated with the motor experiment among all the participants. The reported lines correspond to the average time activations for modes 2 and 3 among all the studied participants (without any additional postprocessing). Finally, the orange lines in \Fig{\ref{fig:Motor-TC}} represent the canonical task-related component expected within each main ROI. We obtained the task-related components using the classical convolutional model with the canonical Hemodynamic Response Function (HRF) \cite{PowFun_2011}. This visualization is crucial because it reflects MVMD's capability to reveal neurophysiological information and how each mode contributes to brain activity. 

Those results showed that IM2 and IM3 effectively capture information related to the expected brain activation patterns within their corresponding ROIs of interest. For instance, mode 2 closely aligns with the block-related activity from the motor cortex ROIs. On the other hand, mode 3, which exhibits a higher central frequency (see Table~\ref{tab:BW-MVMD}), encodes the visual cue associated with the motor task and fixation. 

Interestingly, although mode 3 seems better suited to visual cues, we can see that it also contributes to the motor cortex, which indicates a potential connection between the visual and motor-related areas. This is a very relevant feature because it allows us to explore how the different areas interact at different timescales.

Overall, these findings provide strong evidence that MMD is an effective method for analyzing brain activity and extracting meaningful information. The fact that mode 2 aligns closely with block-related activity in the motor cortex ROIs, while mode 3 captures the visual cues associated with the motor task and fixation, demonstrates the interpretability of the results obtained through MMD. This further supports the validity of MMD in studying the connection between different brain regions and their functional roles.

\subsection{Time activation patterns from the resting state experiment}

\begin{figure}
    \centering
    \includegraphics[width=1.0\columnwidth]{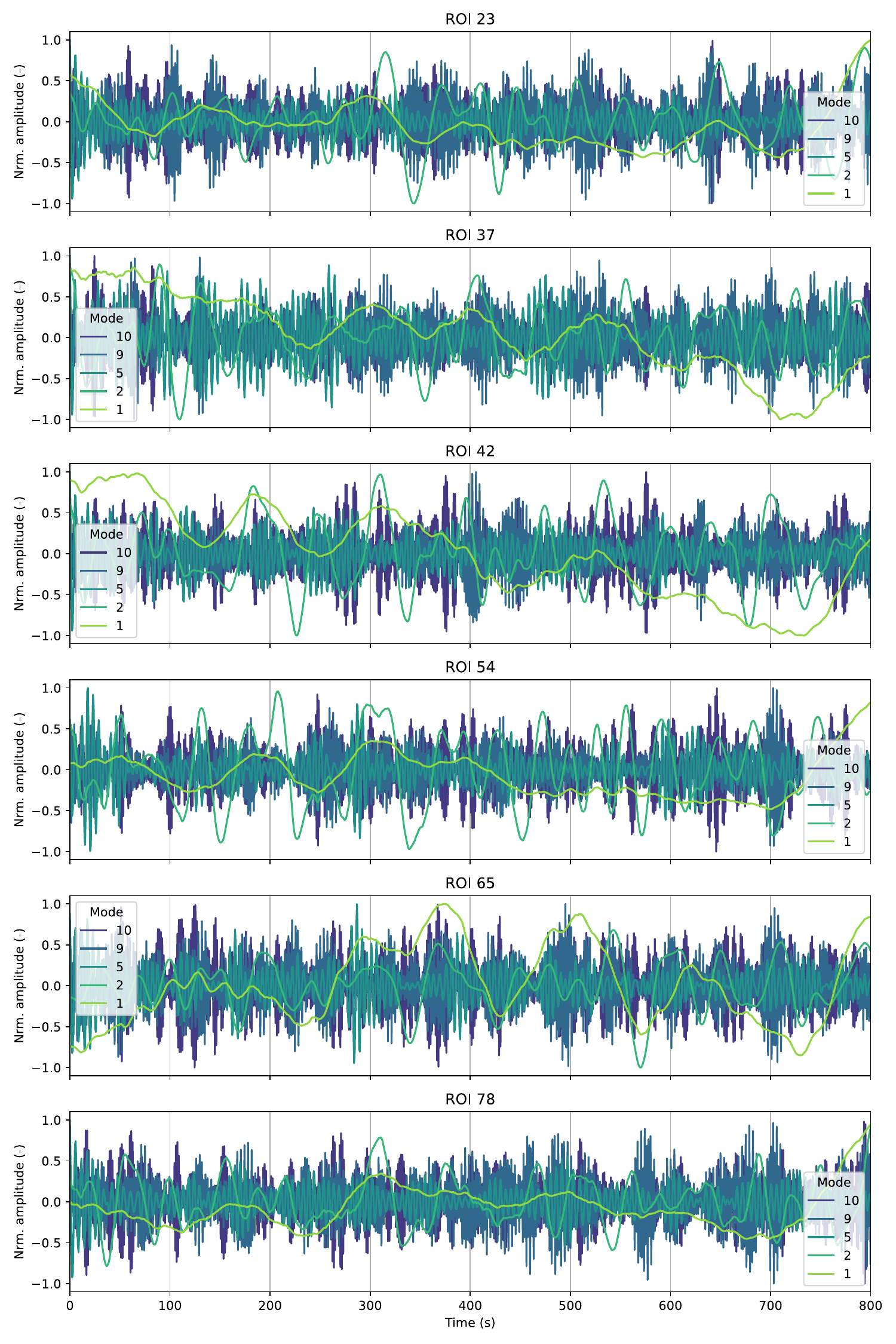}
    \caption{Time activation patterns associated with some IMs among several randomly selected ROIs from a randomly selected individual. The results correspond to the analysis of the resting-state experiment.}
    \label{fig:random-TC}
\end{figure}

For completeness, \Fig{\ref{fig:random-TC}} displays the activation patterns within some randomly selected ROIs. We observed that all the individuals exhibited similar activation patterns in this experiment. Therefore, the figure shows the results for a single randomly selected participant.

\subsection{Intrinsic Modes extraction}
In this study, we used two of the most popular algorithms for MMD decomposition: MVMD \cite{RehMul_2019} and MEMD \cite{RehMul_2009}. Overall, this section aims to illustrate the process of extracting the IMs, understand their physiological meaning, and, at the same time, examine which algorithm provides better results.

\begin{figure*}
    \centering
    \includegraphics[width=\textwidth]{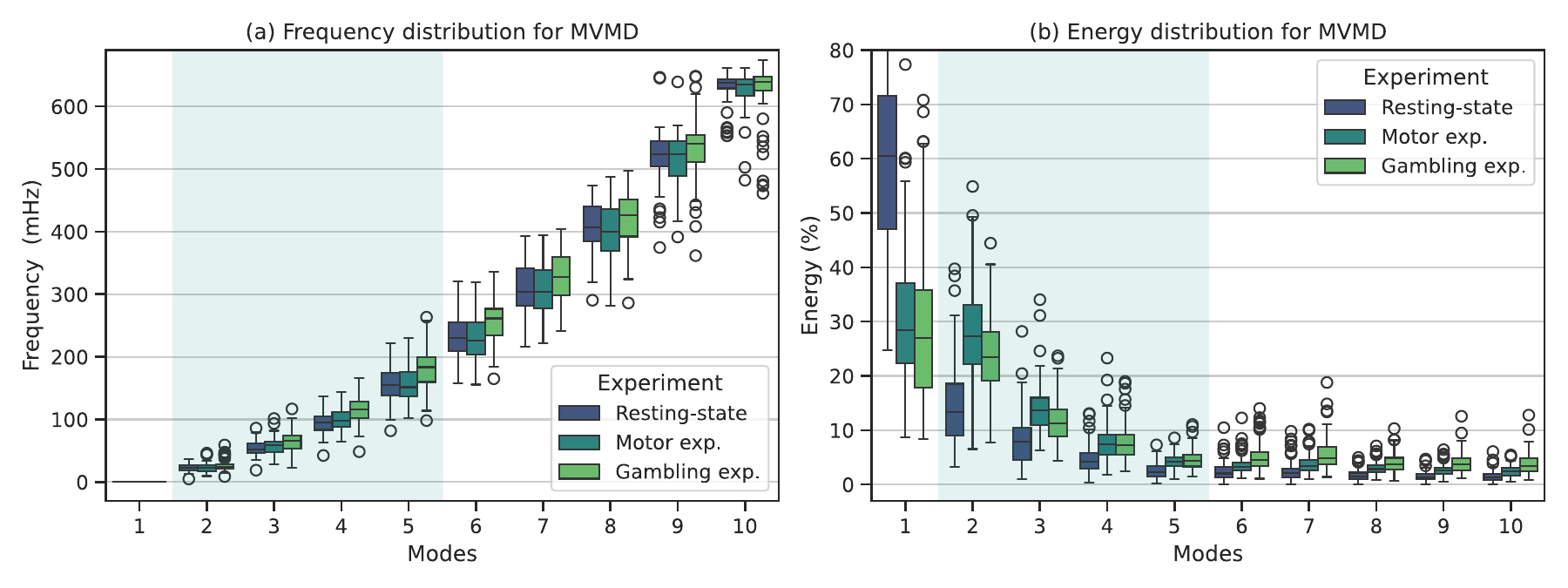}
    \caption{\lFreqEnergy{MVMD}}
    \label{fig:FrquencyEnergy_MVMD}
\end{figure*}

\subsubsection*{IMs extraction using MVMD} \Fig{\ref{fig:FrquencyEnergy_MVMD}} illustrates the frequency (a) and energy (b) distribution of the IMs using the MVMD algorithm for all participants. The first IM exhibits a dominant frequency centered around zero. Similarly, this mode shows a higher relative energy contribution among all the participants. Although we observed a high degree of consensus in frequency and energy among all participants, a closer examination of the actual time activation patterns associated with this mode among individuals and ROIs (see, for example, \Fig{3} in the supplementary material) revealed that those activation patterns vary among participants, which indicates that those low-frequency components are individual-dependent. The main reason for this behavior is that this mode captures trends, low-frequency variations, and motion residuals. Those components vary highly among participants, often spreading through the whole brain. Therefore, they are expected to contribute substantially to the relative energy of the signal.

Modes 2 to 5, highlighted within a shaded area in \Fig{\ref{fig:FrquencyEnergy_MVMD}}, appeared within the neurophysiologically relevant frequency range. This range includes typical brain activity frequencies related to several cognitive and neurophysiological processes \cite{CorFre_2001}. \Tab{\ref{tab:BW-MVMD}} shows the average central frequencies for each IM among participants and their corresponding bandwidths. We found that those results were relatively consistent among participants and experiments.

\begin{table}
    \begin{adjustbox}{width=0.5\textwidth, center}
%
\begin{tabular}{lrrrrrr}
    \toprule
    & \multicolumn{2}{c}{\textbf{Resting state}} & \multicolumn{2}{c}{\textbf{Motor}} & \multicolumn{2}{c}{\textbf{Gambling}}\tabularnewline
     & $\bar{f}$ & BW (mHz) & $\bar{f}$ & BW (mHz) & $\bar{f}$ & BW (mHz)\tabularnewline
    \midrule
    \midrule
    Mode  1 &   0 &  6.8$\pm$  1.0 &   0 & 21.6$\pm$  1.5 &   0 & 26.3$\pm$  1.5 \\
    Mode  2 &  23 & 14.3$\pm$  1.9 &  23 & 36.5$\pm$  4.1 &  26 & 44.9$\pm$  4.3 \\
    Mode  3 &  54 & 15.0$\pm$  2.2 &  56 & 40.5$\pm$  1.9 &  66 & 50.8$\pm$  2.3 \\
    Mode  4 &  96 & 15.9$\pm$  2.7 &  99 & 43.0$\pm$  3.1 & 115 & 53.0$\pm$  2.9 \\
    Mode  5 & 156 & 16.9$\pm$  2.7 & 157 & 47.6$\pm$  3.9 & 182 & 57.4$\pm$  3.6 \\
    Mode  6 & 232 & 16.5$\pm$  2.8 & 231 & 50.6$\pm$  3.6 & 257 & 57.0$\pm$  3.8 \\
    Mode  7 & 310 & 17.2$\pm$  3.1 & 308 & 50.8$\pm$  3.9 & 330 & 57.6$\pm$  4.1 \\
    Mode  8 & 407 & 18.1$\pm$  2.5 & 400 & 54.1$\pm$  4.5 & 419 & 61.6$\pm$  4.4 \\
    Mode  9 & 521 & 18.3$\pm$  2.4 & 514 & 56.6$\pm$  3.9 & 533 & 63.8$\pm$  4.0 \\
    Mode 10 & 632 & 17.4$\pm$  2.3 & 627 & 53.8$\pm$  5.2 & 627 & 56.8$\pm$  7.4 \\
\bottomrule
\end{tabular}
    \end{adjustbox}
    \caption{Average frequency ($\bar{f}$) and their corresponding bandwidth (BW) associated with each IM for the different studied experiments for MVMD.}
    \label{tab:BW-MVMD}
\end{table}

The remaining modes spread across high frequencies. As discussed in Section \ref{sec:NaturalFreq}, signals within this frequency range originate from a mixture of different interfering components. These components include those induced by respiration movements, heartbeat, and cerebrospinal fluid pulsations. Notably, mode 6 --approximately centered at 250 mHz-- captures the primary respiratory-related component, while mode 9, with a central frequency of approximately 520 mHz, aligns well with the first harmonic of the respiratory frequency~\cite{CorFre_2001, YueInt_2019}.

Regarding the relative energy contribution among IMs, illustrated in \Fig{\ref{fig:FrquencyEnergy_MVMD}}, we observed that, in general, modes with lower frequencies exhibited higher energy, while those with higher frequencies had decresaing energy contributions. Specifically, the first three modes contain most of the signal's energy, whereas the high-frequency modes contribute comparatively little. 

Overall, those results obtained from MVMD align with the expected natural behavior of fMRI data as summarized in Section \ref{sec:NaturalFreq}. Therefore, MVMD appears to be a suitable candidate for MMD of fMRI data.
\begin{figure*}
    \centering
    \includegraphics[width=1.0\textwidth]{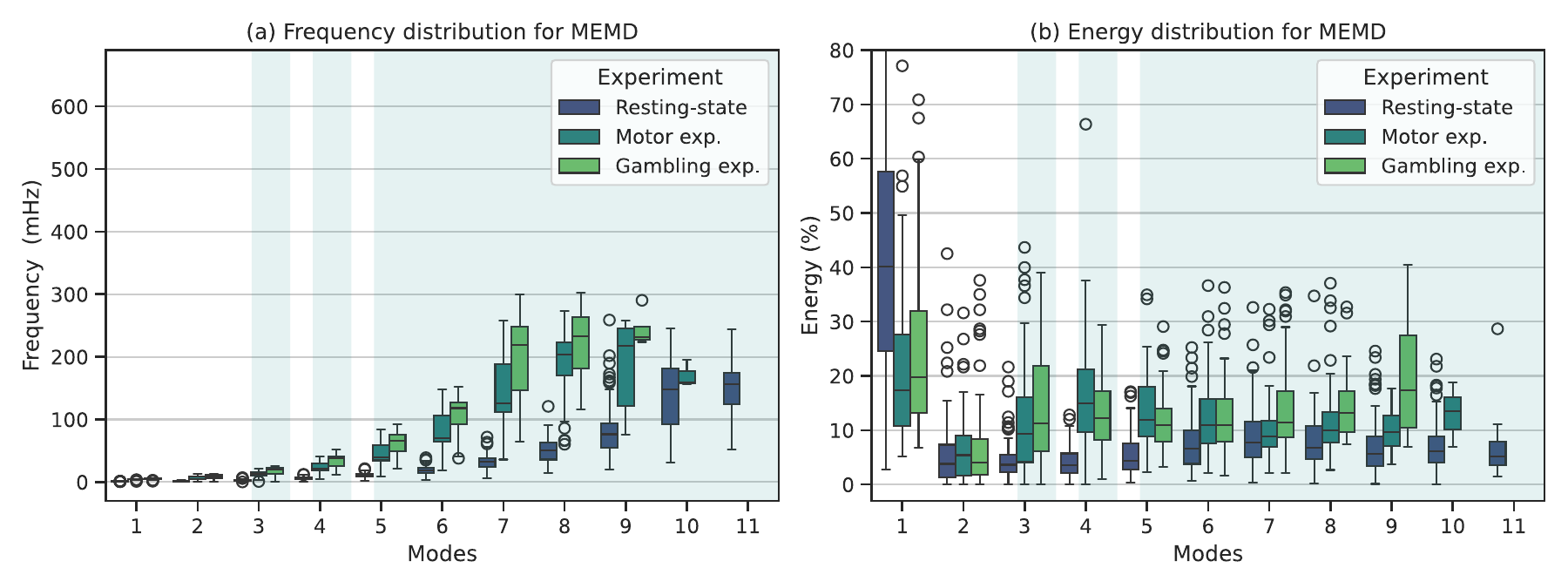}
    \caption{\lFreqEnergy{MEMD}}
    \label{fig:FrquencyEnergy_MEMD}
\end{figure*}

\subsubsection*{IMs extraction using MEMD} \Fig{\ref{fig:FrquencyEnergy_MEMD}} illustrates the frequency (a) and energy (b) distribution of the modes derived from analysis using MEMD among all the participants. However, in contrast to the results of MVMD shown in \Fig{\ref{fig:FrquencyEnergy_MVMD}}, the results from MEMD exhibited a different behavior. Firstly, we observed that the number of intrinsic modes recovered by MEMD varied among individuals and experiments. No doubt, variations in the number of IMs per participant and experiment are expected; each individual may exhibit different active IMs, particularly if they perform different tasks. In this regard, MEMD offers more flexibility than MVMD, which fixes the number of IMs to a specific value. However, the results obtained from MEMD are rather unusual.

We found that most of the identified modes were primarily in the low-frequency range, with nearly all modes from the resting-state experiment falling below 150 mHz, as shown in \Fig{\ref{fig:FrquencyEnergy_MEMD}} Similarly, the results also revealed that many modes within the low-frequency components exhibited large overlapping bandwidths, as shown in \Tab{\ref{tab:BW-MEMD}}. Interestingly, no high-frequency components related to respiratory or cardiac pulsations were observed in both the resting-state and motor experiments.

\begin{table}
    \begin{adjustbox}{width=0.5\textwidth, center}
%
\begin{tabular}{lrrrrrr}
    \toprule
    & \multicolumn{2}{c}{\textbf{Resting state}} & \multicolumn{2}{c}{\textbf{Motor}} & \multicolumn{2}{c}{\textbf{Gambling}}\tabularnewline
     & $\bar{f}$ & BW (mHz) & $\bar{f}$ & BW (mHz) & $\bar{f}$ & BW (mHz)\tabularnewline
    \midrule
    \midrule
    Mode  1 &   1 &  4.6$\pm$  0.1 &   5 & 19.4$\pm$  0.3 &   5 & 24.1$\pm$  0.7 \\
    Mode  2 &   2 &  5.1$\pm$  0.8 &   6 & 20.6$\pm$  4.6 &   9 & 27.3$\pm$  5.4 \\
    Mode  3 &   3 &  6.7$\pm$  1.2 &  12 & 27.6$\pm$  5.0 &  18 & 37.4$\pm$  7.3 \\
    Mode  4 &   6 &  8.5$\pm$  1.1 &  23 & 35.4$\pm$  4.3 &  34 & 46.5$\pm$  4.6 \\
    Mode  5 &  11 &  9.9$\pm$  1.2 &  44 & 41.6$\pm$  5.4 &  63 & 54.4$\pm$  4.9 \\
    Mode  6 &  19 & 12.1$\pm$  1.9 &  81 & 51.9$\pm$  8.3 & 109 & 68.2$\pm$ 10.2 \\
    Mode  7 &  32 & 14.5$\pm$  2.4 & 142 & 60.6$\pm$  9.3 & 202 & 78.3$\pm$ 12.7 \\
    Mode  8 &  51 & 14.4$\pm$  3.3 & 192 & 60.2$\pm$ 10.5 & 222 & 80.2$\pm$ 13.3 \\
    Mode  9 &  82 & 12.8$\pm$  3.4 & 190 & 62.0$\pm$  5.4 & 244 & 76.2$\pm$ 19.2 \\
    Mode 10 & 140 & 12.6$\pm$  2.5 & 170 & 52.8$\pm$  9.6 &     & \\
    Mode 11 & 153 & 12.3$\pm$  2.5 &     &                &     & \\
\bottomrule
\end{tabular}
    \end{adjustbox}
    \caption{Average frequency ($\bar{f}$) and their corresponding bandwidth (BW) associated with each IM for the different studied experiments for MEMD.}
    \label{tab:BW-MEMD}
\end{table}



Finally, when analyzing the energy distribution in \Fig{\ref{fig:FrquencyEnergy_MEMD}}, we observed that the first mode in all the experiments exhibited the most energy contribution, as expected due to the nature of the first mode. However, unlike the results in \Fig{\ref{fig:FrquencyEnergy_MVMD}}, we observed that the energy associated with the rest of the IMs from MEMD exhibited almost a homogeneous distribution.

We must point out that MEMD has no specific parameters to tune, as we discussed in Section 1 of the supplementary material. Consequently, the obtained results cannot be associated with poor parameter selection. We suggest that MEMD led to those unnatural and unreliable results due to the combination of two factors. First, the nature of the interfering components and noise in the data; fMRI data is corrupted with Rician noise \cite{PolHan_2011}, which primarily affects the lower frequencies. Similarly, motion artifacts and trends often have --which usually exhibit relatively high energy-- occur within the lower frequencies. Second, the greedy nature of MEMD algorithm.


In summary, MEMD seems to be driven by low-frequency trends, as most of the components appear dominated by those low trends (see \Tab{\ref{tab:BW-MEMD}}. Therefore, we concluded that MEMD is unsuitable for MMD of fMRI data.

\subsection{Neurophysiological IMs: analysis and interpretation}

Based on the results from the previous section, it is clear that MVMD is an excellent algorithm for MMD of fMRI data. Specifically, MVMD provides a good separation of the IMs across the expected natural frequencies, while MEMD fails to provide interpretable results. This section focuses on the analysis of the results obtained from MVMD among the studied fMRI experiments.

\subsubsection*{Analysis of IMs between fMRI experiments}
Firstly, the frequency and energy distribution for MVMD in \Fig{\ref{fig:FrquencyEnergy_MVMD}} show a similar trend among all the studied experiments. Nonetheless, upon closer examination, we noticed some interesting differences among experiments. For instance, modes 5 and 6 in the gambling experiment exhibited a significantly higher central frequency than the other experiments. Similarly, when examining the energy distribution of these modes, we observed that modes 6 and 7 significantly contributed more to the overall energy of the signal. 

The nature of mode 6 makes this result particularly interesting. These findings suggest that the participant's physiological state during the gambling experiment may have been different from the resting-state and the motor task experiments\footnote{Mode 9 in the gambling experiment, which aligns with the first respiratory harmonic, also showed a higher frequency compared to the other two experiments. Nonetheless, this difference was not statistically significant.}. The higher frequency rate of the respiratory component associated with mode 6 in the gambling experiment and the high energy contributions indicate faster physiological rhythms. This particular physiological state corresponds with a higher level of arousal, which potentially reflects a state of alert and excitement from the gambling experiment, contrasting with the potential physiological states from the other experiments.

Moreover, a closer examination of the energy distribution of low-frequency IMs among the different experiments reveals a significant difference in their energy contributions. In the resting-state experiment, the majority of energy is concentrated in the first mode, representing the dominant and most energetic component. This mode is followed by modes 2 and 3, which also exhibit high energy levels. In contrast, in the task-related experiments, the first mode is less dominant, and neurophysiologically relevant modes (IMs 2-5) consistently show higher energy levels. Modes 2 and 3 capture a substantial portion of the energy, while modes 4 and 5, although showing lower energy contributions, are higher compared to the resting state.

This observed energy distribution between IMs is consistent with the nature of the different explored experiments: during the resting-state experiment, the participants were instructed to remain at rest, with their eyes closed, and think of nothing. Therefore, the brain activity at resting should have a relatively low contribution to the overall signal's energy. By contrast, during task-related experiments, the brain actively participates in a particular task, using more energy as a consequence.

The difference in energy contributions of neurophysiological IMs between resting-state and task-related experiments suggests that the brain is more engaged during task performance compared to a state of rest, which makes intuitive sense. The higher energy levels in neurophysiologically associated modes indicate increased neural activity and processing during the task, highlighting the cognitive demands and involvement required for task completion.

\subsection{Multiscale Functional Connectivity}

Once the IMs have been extracted and identified with neurophysiological information, their FC can be extracted. Nevertheless, we examined the reproducibility of participants' results to ensure relevance. We need to ensure that the IM actually produces consistent results among participants, as extraction of the IM is performed individually. This reproducibility study enabled us to confirm the validity of our proposed approach.

In this regard, note that this step differs from other approaches, such as \cite{AmiQue_2018}, in that it constitutes a necessary step to ensure results consistency. Rather, our approach is robust enough to be shown after IM extraction, without any fine-tuning.

\subsubsection*{Reproducibility of the FC maps among participants}

Reproducibility refers to the ability of methods to consistently detect consistent activity within the expected ROIs among different experimental realizations~\cite{VarGro_2010, MorInf_2020}. In this case, we focus on the reproducibility associated with the FC patterns from MMD among participants. Since all participants are analyzed independently, failing to provide consistent results among participants indicates that the method is unable to capture relevant common information. 

Therefore, we conducted a study to assess the reproducibility of the FC among all participants. This study aimed to demonstrate how different FC patterns behaved across participants. In particular, we studied the individual reproducibility of the static FC patterns associated with each IM for the three considered fMRI experiments.

For completeness, we evaluated the reproducibility of the FC patterns across all IMs. To this end, we calculated the Pearson's correlation obtained from all the possible pairs of comparisons across all participants. We want to emphasize that this step was critical in ensuring the validity of our findings, as it allowed us to confirm that the FC patterns associated with each IM were consistent across all participants, providing a solid foundation for our subsequent analyses.

\begin{figure}
    \hspace{-2mm}\includegraphics[width=1.15\columnwidth]{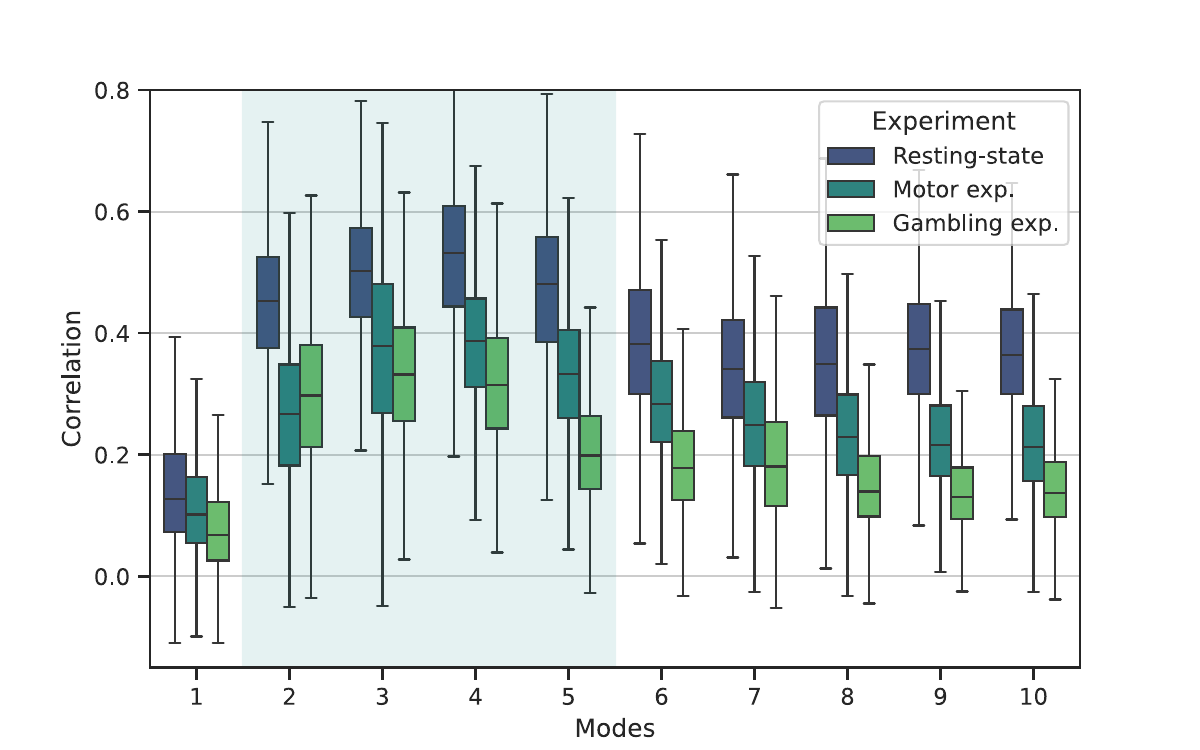}
    \caption{Reproducibility of the static FC patterns associated with each intrinsic mode for the three considered fMRI experiments using MVMD. The boxplots depict the Pearson's correlation values obtained among all the possible individual pair comparisons across all the participants}
    \label{fig:Reliability_MVMD}
\end{figure}

\Fig{\ref{fig:Reliability_MVMD}} illustrates the similarity of the FC associated with each mode using MVMD for the three studied experiments. The boxplots depict the results obtained across all pair comparisons for all participants. 

Upon examination, we observed that all experiments followed a similar trend. Modes within the neurophysiological band exhibited higher similarity than the other modes. Specifically, modes 3 and 4 demonstrate remarkable reproducibility. 

In contrast, mode 1 exhibited the lowest similarity across participants in all experiments. These results indicate that the observed patterns associated with the first mode largely depend on each individual; as we already pointed out, the first mode contains individual trends and motion residuals that depend on each particular individual.

Regarding the remaining modes above the neurophysiological band, even though they are less consistent than modes within the neurophysiological band, they still exhibit a higher similarity to mode 1. This result indicates that the FC associated with these IMs exhibits a certain degree of consistency among individuals despite mostly containing interfering physiological components, as previously discussed.

Although the three studied experiments exhibited a similar trend, a closer examination reveals some interesting differences. First, the resting-state experiment consistently showed increased similarity for all modes, with mode 4 (96 mHz) being the most consistent among all individuals. This may also be related to why resting-state is the dominant approach in FC studies; on top of its simplicity, these results highlight that, in contrast to task-related experiments, activation patterns at different frequencies remain stable, as they seem to produce more reliable FC patterns.

If we focus on the high-frequency modes, modes 8 and 9 maintained a relatively higher similarity than modes 7 or 6. We speculate that this is due to cardiac interfering components at rest. Cardiovascular and cerebrospinal pulsations during rest are more stable and predictable, leading to consistent correlation patterns across individuals in the high-frequency modes \cite{CheRes_2020}.

Conversely, for the motor experiment, we observed that mode 3 (56 mHz) was the most consistent among all the participants, followed closely by mode 4. This result was expected since mode 3 (56 mHz) contains a visual cue common to all participants. The presence of a shared stimulus in this mode likely contributes to its higher consistency among individuals during the motor task. This may also indicate that the visual cue dominates the brain activity among all the participants.

Last but not least, the gambling experiment displayed the lowest similarity across all modes. Nonetheless, we still observed that modes 3 and 4 exhibited the highest consistency among the other modes. Interestingly, we noticed a significant drop in the similarity in the modes at the higher frequencies.

\subsubsection*{Further evaluation of MEMD} 
\begin{figure}
    \hspace{-2mm}\includegraphics[width=1.15\columnwidth]{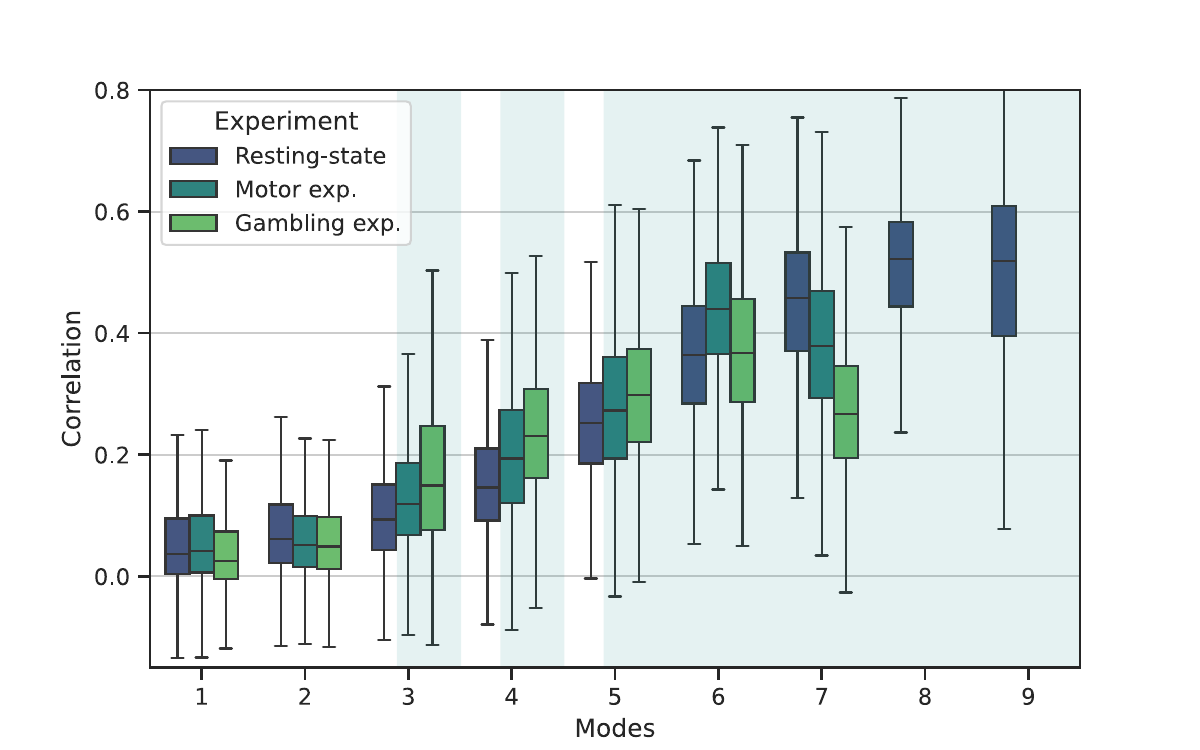}
    \caption{Reproducibility of the static FC patterns associated with each intrinsic mode for the three considered fMRI experiments using MEMD. The boxplots depict the Pearson's correlation values obtained among all the possible individual pair-comparisons across all the participants}
    \label{fig:Reliability_MEMD}
\end{figure}

Although the results in the previous section already evidence the inadecuacy of MEMD to extract relevant information from fMRI data, we decided, for completeness, to further evaluated the reproducibility of the modes obtained using the MEMD algorithm. \Fig{\ref{fig:Reliability_MEMD}} illustrates the obtained results using MEMD. 

As expected, the FC associated ot the IMs from MEMD are not similar among participants. Interestingly, the high-frequency modes exhibited the highest similarity. These results further support the idea that the MEMD algorithm seems to be mostly driven by the noise associated with the high-frequency components rather than the inherent oscillation of the data.

\subsection{Neurophysiological MFC patterns and analysis}

After understanding the frequency and energy distribution of the different IMs, their corresponding FC patterns and the reproducibility among individuals, we focus on the interpretation of the FC patterns associated with the IMs. In this case, we are interested in studying only neurophysiological activation patterns. Similarly, we focus our study on the neurophysiological FC patterns from MVMD, as it provided the best performance.

\begin{figure*}
    \centering
    \includegraphics[width=0.9\textwidth]{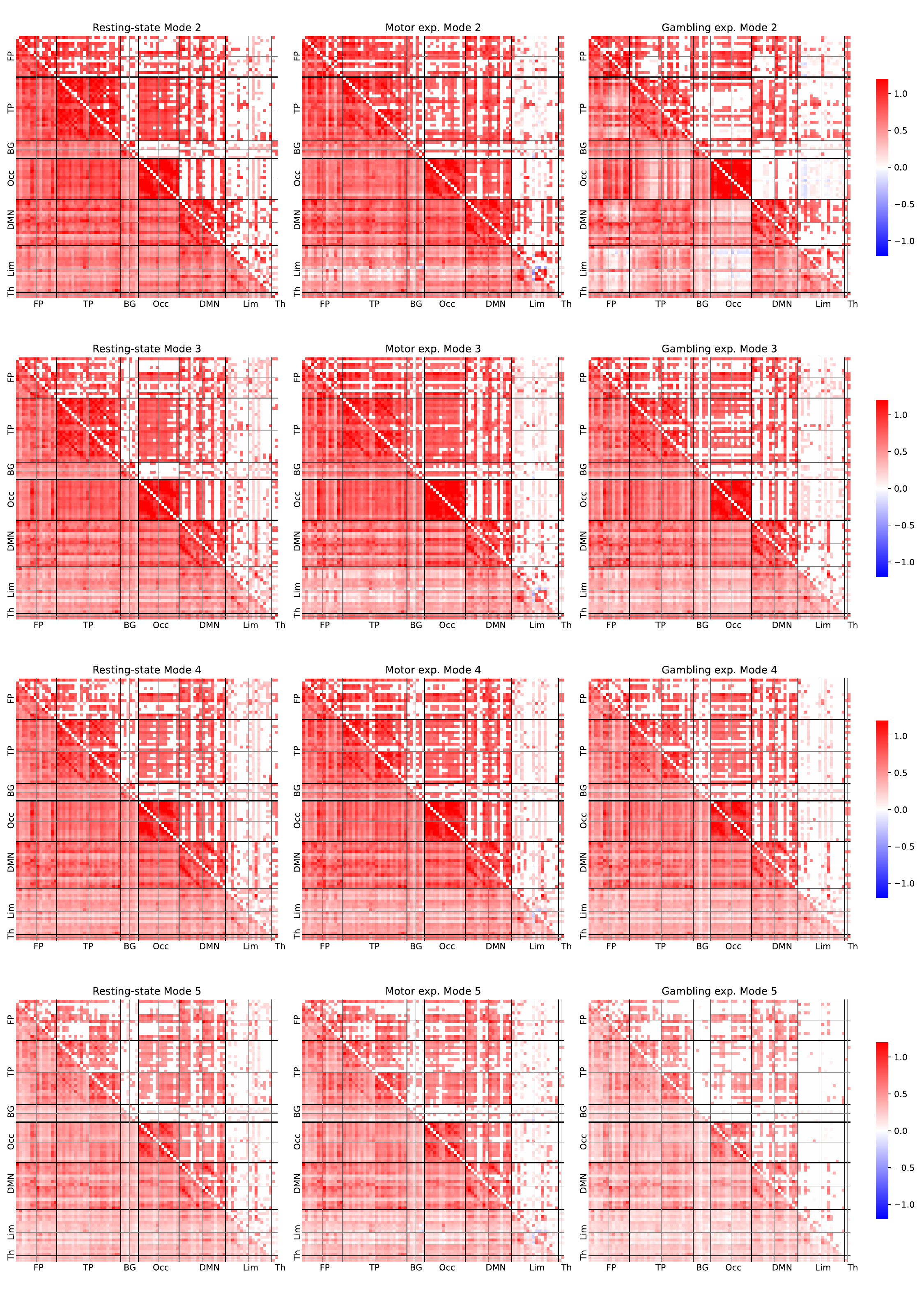}
    \caption{FC patterns for the modes 2, 3, 4 and 5 for the three studied fMRI experiments. The mean FC patterns were estimated by averaging across 100 participants. Person's correlation coefficients were Fisher-Z transformed. The lower diagonal part shows all the averaged correlation coefficients. The upper diagonal only displays significant correlation coefficients compared to the null dataset from a permutation-based t-test corrected with a false positive rate adjusted to $p\leq 0.001$.}
    \label{fig:SFC_patterns}
\end{figure*}

\Fig{\ref{fig:SFC_patterns}} depicts the average FC patterns for each mode. We obtained those FC patterns by averaging them across all participants, as they exhibited a high reproducibility. Each row corresponds to a particular mode, and each column contains different experiments. For all comparisons, we performed a statistical test with respect to a null dataset generated from each particular decomposition by randomly mixing the temporal samples of the IMs. Pearson's correlation coefficients were Fisher-Z transformed. The lower diagonal of each connectivity matrix displays the average correlation coefficients, while the upper diagonal shows only the correlation values that also exhibited significant activation compared to the null data derived from a permutation-based t-test corrected for false discovery rate adjusted to $p<0.001$. For convenience, we arranged the ROIs according to the leading module (left and right), following the order reported in \Tab{\ref{tab:selectedROIs}}.

From a general overview of \Fig{\ref{fig:SFC_patterns}}, we found some similarities among the experiments, particularly within the networks linked to the main modules. We observed some relevant patterns during closer inspection, which we briefly summarize below.

Firstly, the IM 2 was characterized by extensive activity across several brain networks involving numerous ROIs. This connectivity suggests a broad engagement of brain regions in this mode, indicating a complex and integrative role in coordinating diverse neural processes. A notable correlation was observed between the occipital and temporal modules. This connection was considerably weaker in the motor and gambling experiments, suggesting contextual modulation of connectivity. Interestingly, while the temporal cortex and the Default Mode Network (DMN) were disconnected in the resting-state experiment, they were linked during task-related activities. This finding highlights the brain's connectivity dynamics and adaptability across different tasks.

Furthermore, during the gambling experiment in mode 2 FC patterns, pronounced left-hemisphere lateralization was linked with the limbic module, indicating the potential role of emotional processing in gambling tasks \cite{WuDyn_2022}. On the other hand, the DMN and the occipital cortex exhibited a significant correlation during the motor task, which was absent in the different experiments. This suggests a specific involvement of these regions in motor control and visual processing. These observations align with results from FC's temporal dynamics in task-relevant representations during cognitive control \cite{BraTas_2022}.

\section{Limitations and future work}
As with any other study, there are some limitations, which we summarized as follows. 

\begin{itemize}
    \item We analyzed three fMRI datasets with similar imaging protocols. Further experiments with a broader range of datasets and conditions would hepl to fully understand MFC's robustness and applicability.

    \item Exploring dynamic aspects of MFC was beyond the scope of this study. A thorough examination of temporal connectivity dynamics of MFC and how they relate to other dynamic FC approaches could be explored in future work.
    
    \item We focused on a single brain atlas. A comparison with other alternative brain atlas could also provide valuable insights on MFC's performance.
    
    \item Our study focused on the ROI level. Exploring the voxel-level application of MFC could yield fine-grained insights into the brain FC.
    
    \item The current MMD algorithms are applied individually. More advanced algorithms tailored to fMRI data and group-level strategies that integrate information from multiple participants could be explored in future research.
\end{itemize}

\section{Conclusions}
In this study, we proposed a novel methodology for extracting neurophysiological functional information from fMRI data across multiple timescales, referred to as Multiscale Functional Connectivity (MFC), by decomposing fMRI activity into distinct intrinsic modes using Multivariate Mode Decomposition (MMD). To the best of our knowledge, this is the first time such an analysis has been performed on fMRI data. Unlike previous similar studies, e.g., \cite{YueInt_2019}, our proposed approach accounts for the natural nonlinear, nonstationary, and multivariate nature of fMRI while allowing for the particularities of each individual in a data-driven way. As such, MFC provides a more comprehensive view of the underlying fMRI data, which is essential for understanding its dynamics and interactions.

The analysis of three different fMRI experiments revealed that the algorithm Multivariate Variational Mode Decomposition (MVMD) constitutes a suitable candidate for MMD in fMRI data. Our results showed that MVMD provides meaningful results while offering reliable, functional connectivity patterns among participants. The study of these connectivity patterns among different neurophysiological components evidenced the intertwined roles of various brain networks at several timescales. We believe that further adoption of MFC would give more insight into fMRI data and enhance our understanding of brain behavior.

\bibliographystyle{IEEEtran}
\bibliography{Biblio}

\end{document}